# Thermal and mechanical denaturation properties of a DNA model with three sites per nucleotide

Ana-Maria FLORESCU and Marc JOYEUX[(#)]

*Laboratoire Interdisciplinaire de Physique (CNRS UMR 5588),*

*Université Joseph Fourier - Grenoble 1,*

*BP 87, 38402 St Martin d'Hères, France*

**Abstract** - In this paper, we show that the coarse grain model for DNA, which has been proposed recently by Knotts, Rathore, Schwartz and de Pablo (J. Chem. Phys. 126, 084901 (2007)), can be adapted to describe the thermal and mechanical denaturation of long DNA sequences by adjusting slightly the base pairing contribution. The adjusted model leads to (i) critical temperatures for long homogeneous sequences that are in good agreement with both experimental ones and those obtained from statistical models, (ii) a realistic step-like denaturation behaviour for long inhomogeneous sequences, and (iii) critical forces at ambient temperature of the order of 10 pN, close to measured values. The adjusted model furthermore supports the conclusion that the thermal denaturation of long homogeneous sequences corresponds to a first-order phase transition and yields a critical exponent for the critical force equal to $\sigma = 0.70$. This model is both geometrically and energetically realistic, in the sense that the helical structure and the grooves, where most proteins bind, are satisfactorily reproduced, while the energy and force required to break a base pair lie in the expected range. It therefore represents a promising tool for studying the dynamics of DNA-protein specific interactions at an unprecedented detail level.

[(#)] email : Marc.Joyeux@ujf-grenoble.fr

# I - INTRODUCTION

This article deals with the thermal and mechanical denaturation of DNA, that is, the separation of the two strands upon heating [1-6] or application of a force [7-15]. More precisely, we report on our investigations of the denaturation properties of a mesoscopic model, which describes each nucleotide as a set of three interacting sites. The motivations for this study are twofold.

First, all of the models that have been developed up to now to investigate theoretically the thermal and mechanical denaturation of long DNA sequences describe each base pair with a very limited number of degrees of freedom. Actually, in most cases a single degree of freedom is used to represent a base pair. This degree of freedom is usually either the distance between paired bases [16-24] or the state of the base pair (open or closed) [25-32]. A few models additionally take into account the rotation of the bases around the strand axes [33-36] or the bending of the chain [37,38]. At some point, the results and predictions obtained from models with such strongly reduced dimensionality must be compared with those obtained from models that describe more accurately the actual structure of DNA sequences, in order to check their robustness and validity.

The second motivation deals with the investigation of DNA-protein interactions. As for DNA denaturation, an atomistic description of these problems is prohibitively expensive from the point of view of CPU time requirements, so that mechanical mesoscopic models [39-45] are the only alternative to kinetic ones [46-58]. For example, we have recently shown that models where 15 DNA base pairs are represented by a single bead are very useful to investigate *non-specific* DNA-protein interactions [59-61], that is, the alternation of 3-dimensional diffusion in the buffer and 1-dimensional sliding along the DNA sequence, by which the protein scans the DNA sequence while searching for its target [62-70]. Such models are, however, no longer sufficient when it comes to investigate *specific* DNA-protein interactions, that is, the process by which the protein recognizes its target and binds firmly to it. For this purpose, more accurate models are needed, which describe correctly the helical geometry of DNA, as well as the minor and major grooves where most proteins bind. Moreover, since the binding of the protein to its target may involve the opening of one or several base pairs, the thermal and mechanical denaturation properties of these models should also match real ones.

Several mesoscopic models that take the helical structure of DNA and sometimes the grooves into account have been proposed over the past few years [71-80]. These models have however been used uniquely to investigate the denaturation (and in a few cases the renaturation [79,80]) of short oligonucleotides with a few tens of base pairs. It must be emphasized, that this problem differs substantially from the denaturation of long sequences, in the sense that the abrupt denaturation phase transition of long sequences is replaced, in the case of short oligonucleotides, by a much smoother dissociation, due to finite-size effects. The model of Knotts, Rathore, Schwartz and de Pablo (KRSdP) [78] is particularly interesting in the context of the two motivations reported above, because (i) of its precise geometry, and (ii) the authors have already shown that it correctly predicts several physical properties of real sequences, like base specificity, the effect of salt concentration on duplex stability, and the long persistent length of double-stranded DNA. In this model, each nucleotide is mapped onto three interaction sites, so that 9 coordinates are necessary to describe its position. This is about one order of magnitude larger than for most mechanical models aimed at investigating DNA denaturation [16-24], and about two orders of magnitude larger than for "beads and springs" models of DNA and the model we have developed to study DNA-protein interactions [39-45,59-61], but calculations involving a few thousands of base pairs are still affordable with nowadays computers.

The remainder of the paper is consequently organized as follows. The model and the evolution equations are described in Section II. Special attention is paid to the description of the modifications we brought to the original KRSdP model in order to adapt it to the investigation of long sequences. Sections III and IV then respectively describe the thermal and mechanical denaturation properties of this model. In Section V, we discuss the effect on denaturation of two terms of the original KRSdP model, namely the excluded volume term and the extension of base pairing to all complementary base pairs, which are neglected in the main body of this work. We finally conclude in Section VI.

## II - EXPRESSION OF THE MESOSCOPIC MODEL

The KRSdP model used in this work has been described in detail in Section II of Ref. [78]. Nevertheless, we provide here a short description thereof for the sake of completeness and in order to point out clearly the modifications we brought to adapt it to the study of the denaturation of long sequences.

As already mentioned, each nucleotide is mapped onto three interaction sites, namely one site for the phosphate group, one site for the sugar group, and one site for the base. At equilibrium, the phosphate and sugar sites are placed at the center of mass of the respective moities, while the site is place at the N1 position for A and G purine bases and at the N3 position for C and T pyrimidine bases. Reference coordinates for each site determined from the standard B isoform [81] are provided in Table I of Ref. [78]. The reference geometry is illustrated in Fig. 1.

The potential energy $E_{pot}$ of the system includes *six* distinct contributions,

$$E_{pot} = V_{bond} + V_{angle} + V_{dihedral} + V_{stack} + V_{bp} + V_{qq} \; , \tag{II.1}$$

where $V_{bond}$, $V_{angle}$ and $V_{dihedral}$ describe the stretch, bend and torsion contributions, respectively, while $V_{stack}$ denotes the stacking interaction between bases belonging to the same strand, $V_{bp}$ the hydrogen bonding between complementary bases, and $V_{qq}$ the Coulomb interaction between the charged phosphate sites. Note that the original model contains an additional term $V_{ex}$, which describes excluded volume interactions between any two sites that do not interact by means of one of the other six contributions. However, this term is quite expensive from the point of view of CPU time requirements, while it influences only moderately the average quantities we are interested in. It was therefore dropped in the calculations presented in Sections III and IV. Its influence will however be discussed in some detail in Section V. The expressions for the six contributions to $E_{pot}$ are

$$V_{bond} = k_1 \sum_i (d_i - d_i^0)^2 + k_2 \sum_i (d_i - d_i^0)^4$$

$$V_{angle} = \frac{k_\theta}{2} \sum_i (\theta_i - \theta_i^0)^2$$

$$V_{dihedral} = k_\phi \sum_i [1 - \cos(\phi_i - \phi_i^0)]$$

$$V_{stack} = \varepsilon \sum_{i<j} [(\frac{r_{ij}^0}{r_{ij}})^{12} - 2(\frac{r_{ij}^0}{r_{ij}})^6 + 1]$$

$$V_{bp} = \varepsilon_{AT} \sum_{AT \text{ base pairs}} [5(\frac{r_{ij}^0}{r_{ij}})^{12} - 6(\frac{r_{ij}^0}{r_{ij}})^{10} + 1]$$

$$+ \varepsilon_{GC} \sum_{GC \text{ base pairs}} [5(\frac{r_{ij}^0}{r_{ij}})^{12} - 6(\frac{r_{ij}^0}{r_{ij}})^{10} + 1]$$

$$V_{qq} = \frac{e^2}{4\pi\varepsilon_{H_2O}} \sum_{i<j} \frac{\exp(-\frac{r_{ij}}{\kappa_D})}{r_{ij}} \quad . \tag{II.2}$$

In Eq. (II-2), $d_i$ represents the distance between two bound sites, $\theta_i$ the angle formed by three consecutively bound sites, and $\phi_i$ the dihedral angle formed by four consecutively bound sites, all of these sites belonging of course to the same strand. In contrast, $r_{ij}$ stands for the distance between two sites that are not directly bound and that do not necessarily belong to the same strand. $d_i^0$, $\theta_i^0$, $\phi_i^0$, and $r_{ij}^0$ denote the values of these variables for the reference configuration (for numerical values, see Tables III and IV of Ref. [78]). We used $k_1 = 0.26$ kcal mol$^{-1}$ Å$^{-2}$, $k_2 = 26.0$ kcal mol$^{-1}$ Å$^{-4}$, $k_\theta = 104.0$ kcal mol$^{-1}$ rad$^{-2}$, and $k_\phi = 1.04$ kcal mol$^{-1}$, as proposed in Ref. [78] (it is reminded that 1 kcal mol$^{-1}$ ≈ 43.4 meV). We also conformed to Ref. [78] in letting $V_{\text{stack}}$ couple not only bases $i$ and $i+1$ of the same strand, but also bases $i$ and $i+2$, and in using $\varepsilon = 0.26$ kcal mol$^{-1}$.

In contrast, we modified somewhat the base pairing interaction $V_{\text{bp}}$. Indeed, it was assumed in Ref. [78] that $V_{\text{bp}}$ describes hydrogen bonding between any complementary base pair and acts both intra- and interstrand. However, as will be discussed in more detail in Section V, this assumption leads to the unphysical result that many bases form bonds with two complementary bases instead of one at most. Therefore, we assumed in this work that $V_{\text{bp}}$ applies only to bases belonging to the same pair $i$. This modification certainly imposes certain limitations to the model, which will also be discussed further in Section V, but it prevents unphysical multiple base pairings. It also has the consequence that the two DNA strands are less tightly bound and separate at lower temperature compared to the original model. Stated in other words, using $\varepsilon_{\text{AT}} = 2.77$ kcal mol$^{-1}$ and $\varepsilon_{\text{GC}} = 4.16$ kcal mol$^{-1}$ as proposed in [78] in the modified model with the simple pairing scheme results in critical temperatures that are too low. For example, let us consider a 480 base pairs (bp) sequence, hereafter called the ...AAAAA... sequence, one strand of which is composed of adenine bases and the other one of thymine bases. We checked that simulations performed with the parameters of Ref. [78] (that is, $\varepsilon_{\text{AT}} = 2.77$ kcal mol$^{-1}$) and the modified model (that is, assuming that $V_{\text{bp}}$ acts only between bases belonging to the same pair $i$) lead to a critical temperature $T_c \approx 230$ K for this sequence, while it actually lies around 335 K at [Na$^+$] = 50 mM salt concentration. This result is illustrated in Fig. S1 of the Supplementary Material [82]. We therefore adjusted $\varepsilon_{AT}$ and $\varepsilon_{GC}$ to get correct denaturation temperatures for long ...AAAAA... and ...GGGGG... sequences, that is, around 335 and 375 K, respectively. As will be shown below, we found that $\varepsilon_{AT} = 3.90$ kcal mol$^{-1}$ and $\varepsilon_{GC} = 4.37$ kcal mol$^{-1}$ meet this requirement.

At last, in the expression for the Coulomb interaction $V_{qq}$ between the charged phosphate sites, $e$ stands for the charge of the electron, which is actually carried by each phosphate group, $\varepsilon_{H_2O} = 78\varepsilon_0$ for the dielectric constant of water at room temperature, and $\kappa_D = 13.603$ Å for the Debye length at [Na$^+$] = 50 mM salt concentration.

The dynamics of the model was studied by integrating Langevin's equations

$$m_j \frac{d^2 \mathbf{r}_j}{dt^2} = -\nabla E_{pot} - m_j \gamma \frac{d\mathbf{r}_j}{dt} + \sqrt{2 m_j \gamma k_B T} \frac{dW(t)}{dt} \ . \qquad (II.3)$$

where $\mathbf{r}_j$ and $m_j$ denote the position and the mass of site $j$, $\gamma$ is the dissipation coefficient, and $W(t)$ a Wiener process. The first term in the right-hand side of Eq. (II.3) describes internal forces, while the two last terms model the effects of the buffer, namely friction and thermal noise. For numerical purposes, the derivatives in Langevin equations were replaced by finite differences. The position of site $j$ at time step $i+1$, $\mathbf{r}_j^{i+1}$, was consequently obtained from the positions $\mathbf{r}_j^i$ and $\mathbf{r}_j^{i-1}$ at the two previous time steps according to

$$m_j (1 + \frac{\gamma \Delta t}{2}) \mathbf{r}_j^{i+1} = 2 m_j \mathbf{r}_j^i - m_j (1 - \frac{\gamma \Delta t}{2}) \mathbf{r}_j^{i-1} - \nabla E_{pot} \Delta t^2 + \sqrt{2 m_j \gamma k_B T} \, \Delta t^{3/2} w(t) , \qquad (II.4)$$

where $\Delta t$ is the time step and $w(t)$ a normally distributed random function with zero mean and unit variance. We used $\Delta t = 10$ fs and $\gamma = 5$ ns$^{-1}$. Note that in the thermodynamic limit of infinitely long homogeneous sequences, the averages of thermodynamic observables do not depend on the particular value that is assumed for the dissipation coefficient $\gamma$ [83]. It is unfortunately not feasible to investigate with the model above the melting properties of very long sequences, so that most results presented below were obtained with 480 bp long ones. Still, preceding studies dealing with simpler dynamical models suggest that 480 bp sequences are indeed already rather close to the thermodynamic limit (see for example Ref. [84]).

### III - THERMAL DENATURATION

#### A. Critical temperatures and denaturation rates

It has been recognized quite early [85] that the thermal denaturation of DNA homopolymer pairs (that is, sequences whose complementary strands are composed of a single type of bases), is rather straightforward. These homogeneous sequences are characterized by a well-defined melting (or critical) temperature $T_c$. For temperatures $T$ larger than $T_c$, the two strands separate, leading to single-stranded DNA. Denaturation actually starts at the two extremities of the sequence before spreading towards its middle part. This is clearly seen in Fig. 2 of Ref. [36] and can be derived analytically for 1-dimensional models (see for example Eqs. (18) and (19) of Ref. [84]). This point is also illustrated in the left plot of Fig. 2, which shows an instantaneous snapshot of the 480 bp ...AAAAA... sequence with clearly melted extremities.

Typical plots showing the time evolution of the number of open base pairs of the initially closed ...AAAAA... sequence at 340 K, that is, about 5 K above the critical temperature $T_c = 335$ K, are displayed in the lower panel of Fig. 3. Each curve corresponds to a single trajectory integrated for $2.5 \times 10^8$ steps with a different set of random numbers. It is seen that the sequence remains closed for a certain amount of time before the extremities begin to separate. This delay may be quite long and may differ a lot from one trajectory to the other. In contrast, once initiated, denaturation or renaturation (annealing) proceed at comparable rates for all trajectories performed at the same temperature.

Instead of starting from a closed sequence, a sensible method for determining the critical temperature of a homogeneous sequence therefore consists in considering a sequence that is half melted, like the one shown in the left panel of Fig. 2, and in monitoring whether it opens or closes completely at a given temperature $T$. The denaturation (respectively, annealing) rate is then computed by dividing the number of base pairs that opened (respectively, closed) by the time it took for them to open (respectively, close). The critical temperature is the temperature at which the denaturation/annealing rate vanishes. Such a plot of the denaturation rate is provided in Fig. 4 for the 480 bp ...AAAAA... and ...GGGGG... sequences. Negative denaturation rates correspond of course to a positive annealing rate. Each point in this figure was obtained by integrating a single trajectory. The number of required integration steps obviously varied strongly with temperature, typically from $5 \times 10^6$ steps for the ...GGGGG... sequence at 400 K up to $5 \times 10^8$ steps for the ...AAAAA... sequence at 335 K. By using this method, we were able to determine that, for the model described in Section II and the newly adjusted values of $\varepsilon_{AT}$ (3.90 kcal mol$^{-1}$) and $\varepsilon_{GC}$ (4.37 kcal mol$^{-1}$), long ...AAAAA... sequences melt at $T_c = 335.0 \mp 0.5$ K and long ...GGGGG... ones at $T_c = 374.75 \mp 0.25$ K. These melting temperatures are in good agreement with the corresponding values of 336.3 and 373.1 K that are obtained with the statistical model of Ref. [86] by using the parameters of Blossey and Carlon [32] and a $[Na^+] = 50$ mM salt concentration.

Fig. 4 additionally shows that average denaturation and annealing rates increase linearly with $T_c - T$. It must be emphasized that, in contrast with other quantities discussed in this paper, these rates do depend on the particular value that is assumed for the dissipation coefficient $\gamma$ (we used $\gamma = 5$ ns$^{-1}$), but the linear dependence should still be observed for different values of $\gamma$. On the other hand, the fact that the rates reported in Fig. 4 are about two orders of magnitude larger than those estimated from experiments dealing with short oligoribonucleotides [87,88] indicates that one should assume a value of $\gamma$ substantially larger than $\gamma = 5$ ns$^{-1}$ in order for the model to reproduce the true time scales of denaturation/annealing events. For practical purposes (i.e. CPU time requirements), we however used $\gamma = 5$ ns$^{-1}$ for all results reported below.

Still, for the sake of completeness, we independently estimated the correct value for $\gamma$ by computing the evolution with $\gamma$ of the 3-dimensional diffusion coefficient $D_{3D}$ of a 367 bp double-stranded ...AAAAA... sequence at 298 K. More precisely, for each value of $\gamma$, the time evolution of the squared deviation $\|\mathbf{r}(t) - \mathbf{r}(0)\|^2$ of the center of mass of the sequence was averaged over four trajectories integrated for about $10^8$ steps and subsequently fitted against the $<\|\mathbf{r}(t) - \mathbf{r}(0)\|^2> = 6 D_{3D} t$ law. Note that for $\gamma = 10^{13}$ s$^{-1}$ we had to use a time step $\Delta t = 2$ fs instead of $\Delta t = 10$ fs. The results shown in Fig. 5 indicate that, in the range extending from $10^{10}$ to $10^{13}$ s$^{-1}$, $D_{3D}$ decreases as the inverse of $\gamma$ according to $D_{3D} \approx 7.94/\gamma$, where $\gamma$ is expressed in s$^{-1}$ and $D_{3D}$ in m$^2$ s$^{-1}$. Since the experimentally determined value of the diffusion coefficient of a 367 bp sequence is $D_{3D} = 1.58 \times 10^{-11}$ m$^2$ s$^{-1}$ [89], the realistic value for $\gamma$ is close to $5 \times 10^{11}$ s$^{-1}$, that is 500 ns$^{-1}$. This value is 100 times larger than the value assumed in this paper, as already suggested by the predicted denaturation rates discussed just above.

**B. Order and width of the transition**

It can be seen in the bottom plot of Fig. 3 that the process that leads to complete denaturation of a sequence is not uniform. The curves are "noisy", which reflects the fact that, under the antagonistic effects of binding energy and random thermal noise, some parts of the sequence open and close transiently several times before remaining open. The amplitude of oscillations decreases as temperature moves away from the critical temperature. Close to the critical temperature, oscillations are instead very pronounced. For example, the top plot of Fig. 3 displays the time evolution of the number of open base pairs of the initially closed 480 bp ...AAAAA... sequence at 335 K, that is at about the critical temperature. This curve was obtained from a single trajectory integrated for $4.5 \times 10^8$ steps. It is seen that opening/closing oscillations take place at all investigated time scales and can involve several hundreds of base pairs. Obviously, if the length of the sequence is smaller than the amplitude of the oscillations, then the sequence will open even if the temperature is smaller than the critical one. This is the reason, why the denaturation and annealing of short oligonucleotides with a few tens of base pairs are rather smooth transitions, which extend over a substantial temperature range [71-80,84,90,91] and differ drastically from the much narrower melting of long sequences described in the present paper.

The melting transition of (infinitely) long homogeneous sequences is indeed essentially characterized by its order or, more finely, by the critical exponent $\alpha$ that describes the behaviour of the singular parts of the free energy per base pair, $f_{\text{sing}}$, entropy per base pair, $s_{\text{sing}}$, internal energy per base pair, $u_{\text{sing}}$, and specific heat per base pair, $(c_V)_{\text{sing}}$, in the neighbourhood of the critical temperature, according to

$$\begin{aligned}
f_{\text{sing}} &\propto t^{2-\alpha} \\
s_{\text{sing}} &= -\frac{\partial f_{\text{sing}}}{\partial T} \propto t^{1-\alpha} \\
u_{\text{sing}} &= f_{\text{sing}} + T f_{\text{sing}} \propto t^{1-\alpha} \\
(c_V)_{\text{sing}} &= -T \frac{\partial^2 f_{\text{sing}}}{\partial T^2} \propto t^{-\alpha},
\end{aligned} \qquad (\text{III.1})$$

where $t = T/T_c - 1$ denotes the reduced temperature. If $\alpha = 1$, then melting corresponds to a first-order phase transition and $s_{\text{sing}}$ and $u_{\text{sing}}$ display a discontinuity at $T = T_c$. Therefore, the sequence absorbs a large amount of energy when it is heated, but its temperature remains constant till denaturation is not complete. The sequence is actually in a mixed-phase regime, where some parts of the sequence (bubbles) are already melted, while other ones are still closed (double-stranded). This is equivalent to the turbulent mixture of liquid water and vapor bubbles that arises at the boiling point of water. In contrast, if $\alpha < 1$, then melting corresponds to a second-order phase transition. $s_{\text{sing}}$ and $u_{\text{sing}}$ are continuous at $T = T_c$, so that no heat is absorbed at the critical point, like, for example, in the ferromagnetic transition.

The question of the order of the DNA melting transition has been (and it still is) much debated. From the point of view of statistical models, the order of the transition depends on the way the partition function of a loop, and particularly the loop closure exponent $c$, is calculated [32]. When using self-avoiding walks in 3D space, $c$ is numerically estimated to be close to $c \approx 1.75$, which corresponds to a second-order phase transition [92] (the boundary between second- and first-order transitions is $c = 2$ [25,93]). In contrast, when using loops embedded in chains [94,95], which is probably a better approximation, one gets $c \approx 2.15$ [94-96], which corresponds to a first-order transition. Unfortunately, experimental results can be equally well reproduced using sets of parameters with $c \approx 1.75$ and $c \approx 2.15$ [32], so that statistical models cannot help deciding whether the melting transition is first- or second-order.

One-dimensional dynamical models lead to conclusions that are even more ambiguous. Indeed, the easiest way to estimate $\alpha$ consists in considering that the singular part of $c_V$ varies much more rapidly than its non-singular part in the neighbourhood of $T_c$ and, consequently, in deducing $\alpha$ from the slope of log-log plots of the temperature evolution of $c_V$ instead of $(c_V)_{\text{sing}}$ [97-100] (remember also that experimentalists have no means to separate the singular from the non-singular part of the measured specific heat). For realistic one-dimensional DNA models, values of $\alpha$ determined in this way are consistently larger than 1 [90,101,102], which may be interpreted as an indication that DNA denaturation corresponds to a first-order phase transition followed by a crossover to another regime in the last few kelvins below the critical temperature [90]. As far as simulations are concerned, one is instead formally able to separate any quantity into a singular and a non-singular part. For example, it was suggested in Ref. [102] that the free energy per base pair, $f$, may be written as the sum of a singular part, $f_{\text{sing}}$, and a non-singular one, $f_{\text{ns}}$

$$f = f_{\text{ns}} + f_{\text{sing}} \ , \tag{III.2}$$

where $f_{\text{ns}}$ is the average free energy per base pair when the two strands are widely (infinitely) separated, that is, when the sequence is single-stranded. This definition leads to well-behaved quantities. Indeed, the singular part is zero above the dissociation temperature, while the non-singular part behaves smoothly when crossing this temperature. It was furthermore shown that the estimation of $\alpha$ from the slope of log-log plots of the temperature evolution of $f_{\text{sing}}$ according to Eq. (III.1) leads to values that vary between 0.5 and 1, depending on the explicit expression that is assumed for the non-linear part of the stacking interaction in these one-dimensional models [102]. Stated in other words, the order of the melting transition depends on the shape and strength of the stacking interaction.

One might consequently wonder what is the order of the transition predicted by the more realistic model of Section II. While for one-dimensional models all thermodynamics quantities that appear in Eq. (III.1) can be straightforwardly computed by using the Tranfer-Integral operator technique [84,90,101-104], this is, however, unfortunately no longer the case for more complex models. For such models, the only method we could think of consists in extracting the average internal energy per base pair, $u$, from molecular dynamics simulations, Eqs. (II.3)-(II-.4) [36,105]. Practically, $u$ is obtained by averaging $E_{pot}$ over long times and/or many simulations and in dividing the obtained value by the number of base pairs of the sequence. In agreement with Ref. [102], $u$ can then be written as the as the sum of a singular part, $u_{sing}$, and a non-singular one, $u_{ns}$

$$u = u_{ns} + u_{sing} \quad , \tag{III.3}$$

where $u_{ns}$ is the average internal energy per base pair when the two strands are widely (infinitely) separated, that is, when the sequence is single-stranded. From the practical point of view, $u_{ns}$ is easily obtained by launching additional simulations where initial conditions correspond to two widely separated strands instead of double-stranded DNA. $u_{sing}$ is then obtained as the difference between $u$ and $u_{ns}$. Results are shown in Fig. 6 for the 480 bp ...AAAAA... sequence. Each point in this figure was obtained by averaging the internal energy along a single trajectory integrated for $5 \times 10^7$ steps. It is seen in the bottom plot of Fig. 6 that the evolution of $u_{ns}$ with temperature is perfectly linear in all the investigated range (275-353 K), as is also the case for $u$ between 275 and 334 K. (remind that above the critical temperature $T_c = 335$ K the sequence is melted, so that $u = u_{ns}$ and $u_{sing} = 0$). Moreover, the slopes of the two curves are exactly the same, so that the singular part of the internal energy, $u_{sing}$, remains constant between 275 and 334 K. It can indeed be seen in the top plot of Fig. 6 that computed values of $u_{\text{sing}}$ vary by less than 1% in this temperature range. It is admittedly difficult to determine precisely what happens between 334 and 335 K, because of the large fluctuations with long period that occur in this interval (see the top plot of Fig. 3). However, except for this very narrow temperature interval, the step-like behaviour of $u_{\text{sing}}$, which switches abruptly from about -3.30 kcal mol$^{-1}$ below 335 K to 0 above 335 K (see Fig. 6) indicates that for this model the characteristic exponent $α$ is equal to 1, which unambiguously supports the thesis that DNA denaturation corresponds to a first-order phase transition.

Figures 4 and 6 moreover indicate that, in agreement with experimental results, the width of the melting transition predicted by the model of Section II is quite small for long sequences. We indeed estimate that the transition width is smaller than 1 K for the 480 bp ...AAAAA... sequence and than 0.5 K for the 480 bp ...GGGGG... sequence.

### C. Thermal denaturation of inhomogeneous sequences

We next used the model of Section II to compute the thermal denaturation curve of the (inhomogeneous) 1793 bp human β-actin cDNA sequence (NCB entry code NM_001101), which has already been discussed in Refs. [23,36,90,104,106]. It has been known since the early work of Gotoh [6] that the denaturation of sufficiently long inhomogeneous sequences occurs through a series of local openings when temperature is increased and that this multi-step process is clearly reflected in the denaturation curve of inhomogeneous sequences in the 1000-10000 bp range. As can be checked in Fig. 1 of Ref. [104], denaturation of the actin sequence does not start at the extremities of the strands, but rather in narrow AT-rich regions centred around positions $n=1300$, $n=1450$ and $n=1600$, before the sequence abruptly melts for all $n>1100$ at slightly higher temperatures. There is then a plateau of several kelvins before the lower end of the sequence finally melts. The portion of the sequence that melts at the highest temperature is located around the GC-richest region aroung $n$=150. The fingerprints of this multi-step process are clearly seen in the dashed curve of Fig. 7, which shows the temperature evolution of the fraction of open base pairs of the actin sequence obtained from the statistical model of Ref. [86] by using the parameters of Blossey and Carlon [32] and a salt concentration $[Na^+] = 50$ mM. Shown on the same figure are the results obtained with the model of Section II (open circles). It is seen that the melting curves obtained with the two models are in qualitative agreement, although the one obtained with the model proposed in this paper appears to encompass a somewhat broader temperature range.

At this point we should discuss the convergence of the results. The ones presented in Figs. 3 and 4 were obtained by waiting till the sequence completely opened or closed, which is an unambiguous criterion. Moreover, the integration time of the trajectories used for the results in Figs. 5 and 6 (as well as Figs. 9-12 below) was sufficiently long to enable a careful check of their stationarity properties. For example, we checked that splitting each trajectory into two segments and calculating diffusion coefficients, internal energies and mean forces from each segment leads to results that are essentially undistinguishable from those presented in these figures. However, we could not perform a similar check for the actin sequence, because each integration step for a 1793 bp sequence lasts as long as about 15 steps for a 480 bp sequence. Each point in Fig. 7 was consequently obtained from a single trajectory that was integrated for $5\times10^7$ steps and we checked that the fraction of open base pairs did not vary significantly during the last $10^7$ steps. However, it cannot be completely excluded that longer integration times would lead to slightly different results. That is why, we are not absolutely sure of the convergence of the results displayed in this later plot.

## IV - MECHANICAL DENATURATION

While thermal denaturation is achieved by raising the temperature of a sequence up to (or above) its critical temperature $T_c$, mechanical denaturation may instead be achieved at temperatures $T < T_c$ by pulling on the extremities of the strands. Mechanical DNA unzipping experiments are usually performed either at constant pulling rate [7,8] or at constant force [12-15]. In the former case, the externally applied force is adjusted to compensate for the action of internal restoring forces exerted by the two strands [7,8]. At ambient temperature, the typical force that must be exerted to keep the two strands open lies in the range 10-15 pN [7-15].

From a practical point of view, constant rate experiments are conveniently modelled by separating slowly the two phosphate groups linked to the first base pair of the sequence and computing the average of the projection along the separation direction of internal forces acting on these phosphate groups. The result of such a simulation performed for the 480 bp ...AAAAA... sequence at 280 K is shown in Fig. 8. Phosphate groups were separated at the speed of 2 cm s$^{-1}$ and each point represents the value of the projected force $F$ averaged over 0.25 ns, that is 25000 integration steps. The complete curve shown in Fig. 8 corresponds to $10^8$ integration steps. Comparison of Fig. 8 with Figs. 5 and 6 of Ref. [107] indicates that separation-force plots obtained with the model of Section II are qualitatively similar to those obtained with one-dimensional models. They essentially consist of a rather large force barrier at short distances [108] followed by a plateau. It has been known since the pioneering work of Ref. [7] that this plateau is flat in the case of an homogeneous sequence, but that it instead displays fluctuations proportional to the local AT/GC concentration when an inhomogenous sequence is being unzipped. We checked that use of the more realistic value $\gamma = 500$ ns$^{-1}$ instead of $\gamma = 5$ ns$^{-1}$ leads to a plot that does not differ significantly from Fig. 8.

The maximum force along the force-displacement curve is known as the threshold force $F_{\text{thr}}$. It may be simply estimated from the maximum of the $F = F(d)$ curve, as illustrated in the inset of Fig. 8. On the other hand, the average asymptotic force at large displacements $d$ represents the critical force $F_c$ that must be exerted on the widely separated extremities of a long sequence to prevent it from zipping. From a practical point of view, $F_c$ was obtained by averaging $F$ over all separations $d$ comprised between 100 and 200 Å. Fig. 8 indicates that the value of $F_c$ obtained in this way is close to 13 pN for the ...AAAAA... sequence at 280 K, in fair agreement with experimentally measured values. We checked that the computed value of $F_c$ does not vary when the pulling rate is divided by a factor two and the strands are separated at the speed of 1 cm s$^{-1}$ instead of 2 cm s$^{-1}$.

Figure 9 further shows the computed temperature evolution of the threshold force $F_{thr}$ and the critical force $F_c$ for the 480 bp ...AAAAA... sequence. Each point in this figure was obtained from a curve similar to that in Fig. 8 but integrated at a different temperature. It appears that $F_{thr}$ decreases slowly (with a slope of about -1.5 pN K$^{-1}$) in the investigated temperature range, from 250 pN at 275 K down to 150 pN at the melting temperature $T_c = 335$ K. In contrast, $F_c$ obviously decreases down to zero at the critical temperature, since at $T = T_c$ thermal denaturation occurs spontaneously, i.e. without application of an external force. The bottom plot of Fig. 9 furthermore indicates that computed values of $F_c$ follow rather precisely a power law of the form $F_c = 0.77(T_c - T)^{0.70}$. Stated in other words, the critical exponent of the critical force is equal to $\sigma = 0.70$.

At this point, it is worth mentioning that, for the several one-dimensional models that were considered, the critical exponents of the specific heat, $\alpha$, and the critical force, $\sigma$, were shown to be related through the linear relation [102]

$$2 - \alpha = 2\sigma .$$
(IV.1)

Since $\alpha = 1$ for the model used here (see Section IIIB), Eq. (IV.1) would suggest that $\sigma = \frac{1}{2}$ instead of $\sigma = 0.70$. Conclusion therefore is that the scaling law of Eq. (IV.1) no longer holds for the more complex model analysed in this paper. On the other hand, it has also been shown that, for the Poland-Scheraga model where self-avoiding interactions are taken into account, the critical force scales like $F_c \propto (T_c - T)^\nu$, where $\nu$ is the correlation length exponent of a self-avoiding random walk [109]. Numerically, $\nu$ is close to 0.588 for a 3-dimensional walk. This value again differs somewhat from the computed exponent, $\sigma = 0.70$.

Last but not least, it should be reminded that experiments actually point out that the phase diagram of mechanical denaturation in the temperature-force plane may be more complex than a simple law of the form $F_c \propto (T_c - T)^\sigma$ anyway [13].

## V - EFFECT OF EXCLUDED VOLUME AND GENERALIZED BASE PAIRING

In this section, we discuss the effect of two terms, which are included in the original KRSdP model [78] but were discarded in the computations that led to the results presented in Sections III and IV, namely the excluded volume interaction term and the generalization of base pairing to all complementary base pairs.

The original KRSdP model indeed contains a seventh term in addition to the six terms of Eq. (II.2). This additional term is an excluded volume interaction term, which insures that sites $i$ and $j$ do not overlap spatially. It writes

$$V_{\text{ex}} = \varepsilon \sum_{i<j} H(r_{ij}^{\text{cut}} - r_{ij}) [(\frac{r_{ij}^{\text{cut}}}{r_{ij}})^{12} - 2(\frac{r_{ij}^{\text{cut}}}{r_{ij}})^6 + 1], \qquad (V.1)$$

where $H(x)$ is the Heaviside step function, which is equal to 0 if $x < 0$ and to 1 if $x \geq 0$. $r_{ij}^{\text{cut}}$ is the threshold at which $V_{\text{ex}}$ starts repelling site $i$ away from site $j$. It was assumed in Ref.

[78] that $r_{ij}^{cut} = 1.00$ Å if sites $i$ and $j$ are two mismatched bases and $r_{ij}^{cut} = 6.86$ Å if sites $i$ and $j$ are any other pair of sites. Included in the sum of Eq. (V.1) are all couples of sites $i$ and $j$, which do not belong to the same strand and to the same or nearest-neighbour base pairs and, additionally, which do not interact through the pairing, stacking or electrostatic interactions (since these later interactions also have a repulsive core).

The essential reason, why $V_{ex}$ was excluded from the calculations discussed in Sections III and IV, is that this approximation reduces the required CPU time by a factor 2, while results are expected to remain essentially unchanged at physiological temperatures. Indeed, at these temperatures the sequence is almost entirely double-stranded, so that neighbouring sites are not likely to overlap spatially, while electrostatic repulsion between charged phosphate sites insures that widely separated portions of the sequence do not cross. This is, however, no longer the case close to the critical temperature, where large bubbles are observed. We consequently discuss below to which extent the excluded volume interaction term affects the results presented in the previous sections.

Fig. 10 shows the computed temperature evolution of the denaturation/annealing rate for a 480 bp homogeneous sequence with pairing energy of 4.37 kcal mol$^{-1}$ (like ...GGGGG... sequence in Sections III and IV) when $V_{ex}$ is taken into account. It may be seen by comparing this plot to Fig. 4 that $V_{ex}$ has two main effects. First, the critical temperature is displaced by about 40 K to lower temperatures, that is, from 375 K down to 335 K. Moreover, denaturation (but not annealing) rates appear to saturate at rather low values, smaller than 0.5 base pairs per nanosecond, when $V_{ex}$ is taken into account (compare Fig. 10 with Fig. 4 and Supplementary figure S2 [110]). This is most probably due to the fact that denaturation pathways become more complex when excluded volume repulsion is taken into account. This is a point, which would certainly deserve more attention.

Finally, we checked that homogenous sequences with base pairing energies of 4.68 kcal mol$^{-1}$ denaturate at about 360 K when $V_{ex}$ is taken into account (see Fig. 10). Linear extrapolation consequently indicates that the values of pairing energies that lead to correct melting temperatures at $[Na^+] = 50$ mM salt concentration are $\varepsilon_{AT} = 4.37$ kcal mol$^{-1}$ and $\varepsilon_{GC} = 4.87$ kcal mol$^{-1}$ when $V_{ex}$ is taken into account, instead of $\varepsilon_{AT} = 3.90$ kcal mol$^{-1}$ and $\varepsilon_{GC} = 4.37$ kcal mol$^{-1}$ when it is neglected.

In contrast, we checked that taking $V_{ex}$ into account does not modify the value of the 3-dimensional diffusion coefficient of a sequence (see Fig. 5). Therefore, the realistic value for $\gamma$ is still close to 500 ns$^{-1}$. Similarly, we checked that the temperature evolution of the singular part of the internal energy, $u_{sing}$, still displays a clear step at the critical temperature when $V_{ex}$ is taken into account (see Fig. 11). This indicates that $V_{ex}$ does not change the predicted order of the denaturation transition, which remains first-order.

At last, let us mention that we were unable to check whether the excluded volume term affects the critical exponent of the critical force. The reason is that pulling rates of the order of 2 cm s$^{-1}$, like that used in Section IV, lead to extremely large computed forces when $V_{ex}$ is taken into account. As for denaturation rates, the reason is probably that pathways for denaturation become more complex. The pulling rate should consequently be sufficiently low to provide the system with sufficient time to follow these complex pathways. From a practical point of view, this however means that simulations become much too long to be feasible.

The last difference between the model of Section II and the original KRSdP model consists in the fact that it was assumed in Ref. [78] that the base pairing term $V_{bp}$ "describes hydrogen bonding between any complementary base pair and acts both intra- and interstrand" [78], while we instead assumed up to now that $V_{bp}$ connects only bases belonging to the same pair. We launched several simulations to check the influence of such a generalized base pairing scheme. The essential result is that generalized base pairing increases the melting temperature of a sequence by several tens of kelvins. For example, simulations showed that a 480 bp ...GGGGG... sequence is still zipped at 390 K when both the excluded volume term and generalized base pairing are taken into account, while it melts at about 335 K when our simpler base pairing scheme is used. The reason for this discrepancy can be understood by computing the distribution of the number $n$ of bonds that is formed by each base. For a given base $i$, $n$ can be estimated at a given time $t$ according to

$$n = \sum_j H(-5(\frac{r_{ij}^0}{r_{ij}})^{12} + 6(\frac{r_{ij}^0}{r_{ij}})^{10})[-5(\frac{r_{ij}^0}{r_{ij}})^{12} + 6(\frac{r_{ij}^0}{r_{ij}})^{10}], \qquad (V.2)$$

where $H(x)$ is again the Heaviside step function and the sum runs over all bases $j$ that are connected to base $i$ through $V_{bp}$. For the simple pairing scheme, there is of course only one base $j$ that may contribute to the sum, while for the generalized base pairing scheme any base $j$ that is complementary to base $i$ can potentially contribute. Moreover, only attractive interactions contribute to $n$ (this is the role of the Heaviside step function), with a weight that varies between 1 (when the distance $r_{ij}$ between the two bases is equal to the equilibrium distance $r_{ij}^0$) and 0 (when $r_{ij}$ becomes very large or, conversely, sufficiently small for the interaction to become repulsive).

The normalized distributions $P(n)$ are shown in Fig. 12 for a 480 bp homogeneous sequence with a pairing energy of 4.37 kcal mol$^{-1}$ (like ...GGGGG... sequence in Sections III and IV). Temperature is 330 K for the simple base pairing scheme (dashed blue line) and 390 K for the generalized base pairing scheme (solid red line). Each distribution was obtained by computing $n$ according to Eq. (V.2) at each step and for each base pair of the sequence along a single trajectory integrated for $1.5 \times 10^6$ steps. Not surprisingly, the $P(n)$ distribution exhibits a single peak culminating at $n = 1$ for the simple base pairing scheme. Less expected is the fact that, for the generalized pairing scheme, $P(n)$ additionally exhibits a second peak culminating at $n=2$, which is much higher than the peak at $n=1$. This indicates that the strands deform in such a way that many bases are able to form two pairing bonds with successive bases of the opposite strand. These double bonds are of course more difficult to break than a single one, which explains why the melting temperature is higher for the generalized base pairing scheme than for the simple one.

Needless to say that such pairing of a base with two complementary ones is not physically correct. In real DNA, a base can pair with at most one complementary base, although this base is not necessarily the complementary one located at the same position on the opposite strand, as is assumed by the simple base pairing scheme. We built several other pairing schemes based on the expression of $V_{bp}$ in Eq. (II.2), where each base is allowed to pair with at most one complementary base, typically the nearest one. However, these schemes turned out to be numerically unstable, because rapid switching of the bonds due to the nearest neighbour criterion led to an artificial increase in kinetic energy and hence of the temperature of the sequence. Following these unsuccessful trials, our feeling is that any improvement in the description of base pairing should take the relative orientations of the sugar-base bonds into account, which would "naturally" select the correct complementary base, if any, to pair with. We leave such an improvement for an eventual future work. Here we instead choose to use the simple base pairing scheme, although we are aware of the limitations it brings to the model. The main limitation is probably that transient pairing of partially matched parts of the sequence cannot take place close to the critical temperature, but we are convinced that this neglect has little consequence on the results presented above. As already emphasized in Section II, use of the simple base pairing scheme instead of the generalized one also has the practical consequence that the values of base pairing energies must be increased compared to the original KRSdP model [78].

## VI - CONCLUSION

In this paper, we have shown that the KRSdP model can be adapted to describe the thermal and mechanical denaturation of long homogeneous and inhomogeneous DNA sequences. This was achieved by using the simple base pairing scheme instead of the generalized originally proposed in Ref. [78] and in consequently refining the two parameters for base pairing interactions, namely $\varepsilon_{AT} = 3.90$ kcal mol$^{-1}$ and $\varepsilon_{GC} = 4.37$ kcal mol$^{-1}$ when the excluded volume interaction term is neglected, and $\varepsilon_{AT} = 4.37$ kcal mol$^{-1}$ and $\varepsilon_{GC} = 4.87$ kcal mol$^{-1}$ when $V_{ex}$ is taken into account. These modifications lead to

(*) critical temperatures for long ...AAAAA... and ...GGGGG... sequences, namely 335 and 375 K, which are in good agreement with both experimental ones and those obtained from statistical models,

(*) a realistic step-like denaturation behaviour for inhomogeneous sequences,

(*) a critical force at ambient temperature of the order of 10 pN, which is also close to measured values.

The modified model furthermore supports the thesis that the thermal denaturation of long homogeneous sequences corresponds to a first-order phase transition. In contrast, it yields a critical exponent for the critical force equal to $\sigma = 0.70$, which suggests that the scaling law relating the characteristic exponents for specific heat and critical force that was derived for one-dimensional models no longer holds for this more complex model.

The KRSdP model with refined base pairing energies therefore represents a good compromise for studying the dynamics of DNA-protein specific interactions at an unprecedented detail level. It is indeed both geometrically and energetically realistic, in the sense that (i) the helical structure and the grooves, where most proteins bind, are satisfactorily reproduced, and (ii) the energy and force required to break a base pair lie in the expected range. We are therefore confident that the combination of this model with a mesoscopic model describing proteins at a comparable level of accuracy will provide new and important insights into this fundamental but very complex problem.

**FIGURE CAPTIONS**

**Figure 1** : Schematic representation of the model, showing the positions at equilibrium of the various sites for a short stretch of the actin sequence discussed in Section III. Letter P indicates a phosphate group (red), S a sugar group (green), and B a base (blue).

**Figure 2** : Snapshots of the denaturation of the 480 bp ...AAAAA... homogeneous sequence (left) and the 1793 bp inhomogeneous actin sequence (NCB entry code NM_001101) (right).

**Figure 3** : Time evolution of the number of open base pairs for the 480 bp ...AAAAA... sequences at 335 K (top plot) and 340 K (bottom plot). In the bottom plot, each curve corresponds to a single trajectory obtained with a different set of random numbers. A given base pair is considered as open if its pairing energy is smaller than $\varepsilon_{AT}/10$. The critical temperature of this sequence is very close to 335 K.

**Figure 4** : Denaturation/annealing rates for 480 bp ...AAAAA... (blue ×) and ...GGGGG... (red +) sequences as a function of temperature, expressed in base pairs per nanosecond. Positive (respectively, negative) rates correspond to denaturation (respectively, annealing). The vertical dash-dotted lines indicate the positions of the critical temperatures (335.0 and 374.75 K, respectively) where the rates are zero. These rates were obtained by starting simulations with half-open sequences like the one shown in the left side of Fig. 2 and in checking how long it takes for these sequences to open or close completely.

**Figure 5** : Evolution with $\gamma$ of the 3-dimensional diffusion coefficient $D_{3D}$ of a 367 bp double-stranded ...AAAAA... sequence at 298 K. $\gamma$ is expressed in s$^{-1}$ and $D_{3D}$ in m$^2$ s$^{-1}$. Red

circles (respectively, blue squares) denote results obtained without (respectively, with) the excluded volume term discussed in Section V. Values of $D_{3D}$ were computed by adjusting the mean squared deviation of the center of mass of the sequence against the $<\|\mathbf{r}(t)-\mathbf{r}(0)\|^2> = 6 D_{3D} t$ law. The solid line shows the $D_{3D} \approx 7.94/\gamma$ law, which best interpolates between the computed points. The horizontal dot-dashed line shows the experimentally determined value of the diffusion coefficient of a 367 bp sequence, that is $D_{3D} = 1.58 \times 10^{-11}$ m$^2$ s$^{-1}$ [89].

**Figure 6** : Temperature evolution of $u$ and $u_{ns}$ (bottom plot) and $-u_{sing}$ (top plot) for the 480 bp ...AAAAA... sequence. $u$ (respectively, $u_{sing}$) denotes the average internal energy per base pair for the double-stranded (respectively, single-stranded) sequence, while $u_{sing} = u - u_{ns}$. The vertical dash-dotted lines indicate the position of the critical temperature $T_c = 335$ K. Remind that $u = u_{ns}$ and $u_{sing} = 0$ above $T_c$.

**Figure 7** : Denaturation curve of the 1793 bp human β-actin cDNA sequence (NCB entry code NM_001101) obtained from the statistical model of Ref. [86] by using the parameters of Blossey and Carlon [32] and a salt concentration $[\text{Na}^+] = 50$ mM (dashed line), and from molecular dynamics simulations performed with the model described in Section II (empty circles). A base pair is considered as open if its pairing energy is on average smaller than $\varepsilon_{AT}/10$ or $\varepsilon_{GC}/10$.

**Figure 8** : Plot, as a function of $d$, of the average force $F$ that must be exerted to maintain the two phosphate groups linked to the first base pair of the 480 bp ...AAAAA... sequence separated by a distance $d$ at $T = 280$ K. $d = 0$ corresponds to the equilibrium distance of the two phosphates. $F$ is the force projected on the separation vector and averaged over 0.25 ns. The insert shows in more detail the large force barrier $F_{thr}$ that occurs at short distances. Experimentally measured critical forces $F_c$ correspond to the asymptotic value of $F$ for large values of $d$, that is, to the average force that must be exerted on the widely separated extremities of long sequences to prevent them from closing.

**Figure 9** : Temperature evolution of the threshold force $F_{thr}$ (top plot) and the critical force $F_c$ (bottom plot) for the 480 bp ...AAAAA... sequence. $F_{thr}$ was obtained as the maximum of $F$ in force-separation curves like that shown in Fig. 8 (averaging time for each point is 0.25 ns). These values are represented as brown × in the top plot. $F_c$ was obtained from force-separation curves as the average of $F$ for all separations comprised between 100 and 200 Å (averaging time for each point is consequently 0.5 µs). These values are represented as blue crosses in the bottom plot. Solid lines show the best adjustment of a linear law (top plot) and a power law (bottom plot) against computed forces.

**Figure 10** : Denaturation/annealing rates for 480 bp homogeneous sequences with pairing energies of 4.37 kcal mol$^{-1}$ (blue ×) and 4.68 kcal mol$^{-1}$ (red +) when the excluded volume term $V_{ex}$ is taken into account. Positive (respectively, negative) rates correspond to denaturation (respectively, annealing). The vertical dash-dotted lines indicate the positions of the critical temperatures (about 335 and 360 K, respectively). These rates were obtained by starting simulations with half-open sequences, like the one shown in the left side of Fig. 2, and in checking how long it takes for these sequences to open or close completely.

**Figure 11** : Same as Fig. 6, except that the excluded volume interaction term of Eq. (V.1) is taken into account in the expression of the Hamiltonian and the investigated sequence has a base pairing energy of 4.37 kcal mol$^{-1}$, as ...GGGGG... sequences in Sections III and IV. Critical temperature is close to 335 K.

**Figure 12** : Distribution of the number of bonds $n$ formed by each base for a 480 bp homogeneous sequence with a pairing energy of 4.37 kcal mol$^{-1}$, as ...GGGGG... sequences in Sections III and IV. Temperature is $T = 330$ K for the simple base pairing scheme (dashed blue line) and $T = 390$ K for the generalized base pairing scheme (solid red line). Unpaired bases are not taken into account in the statistics.

**FIGURE 1**

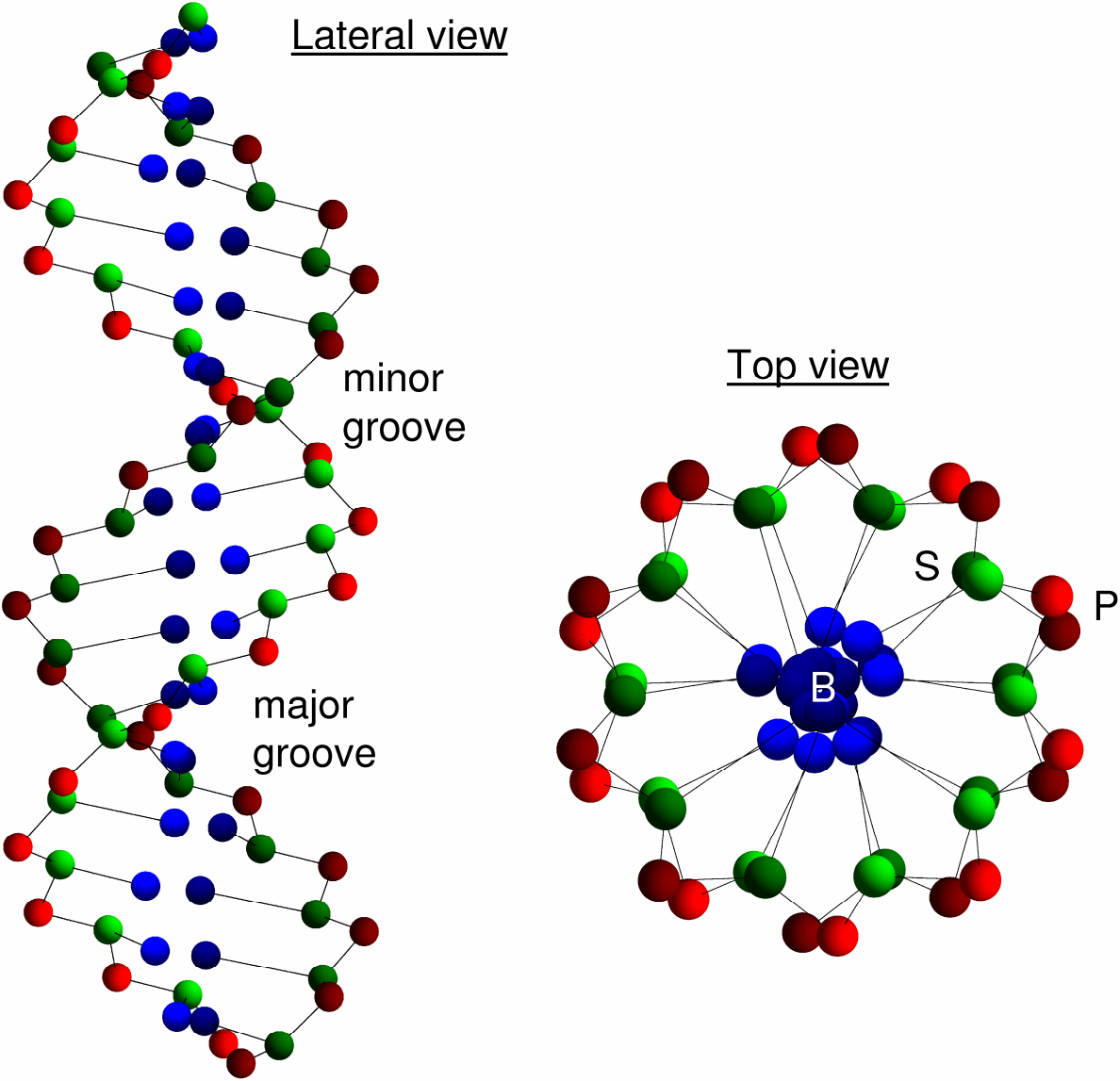

**FIGURE 2**

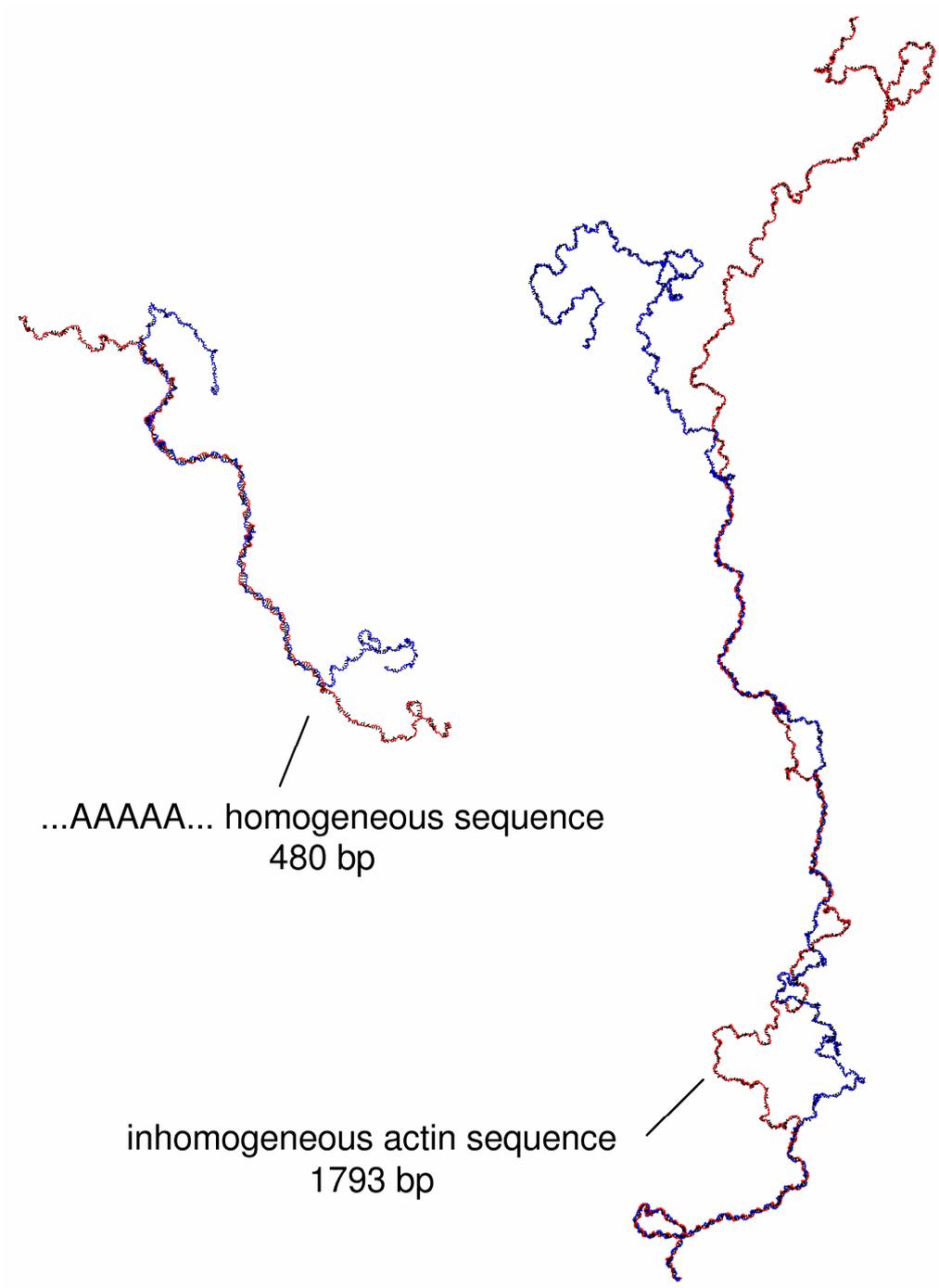

...AAAAA... homogeneous sequence
480 bp

inhomogeneous actin sequence
1793 bp

**FIGURE 3**

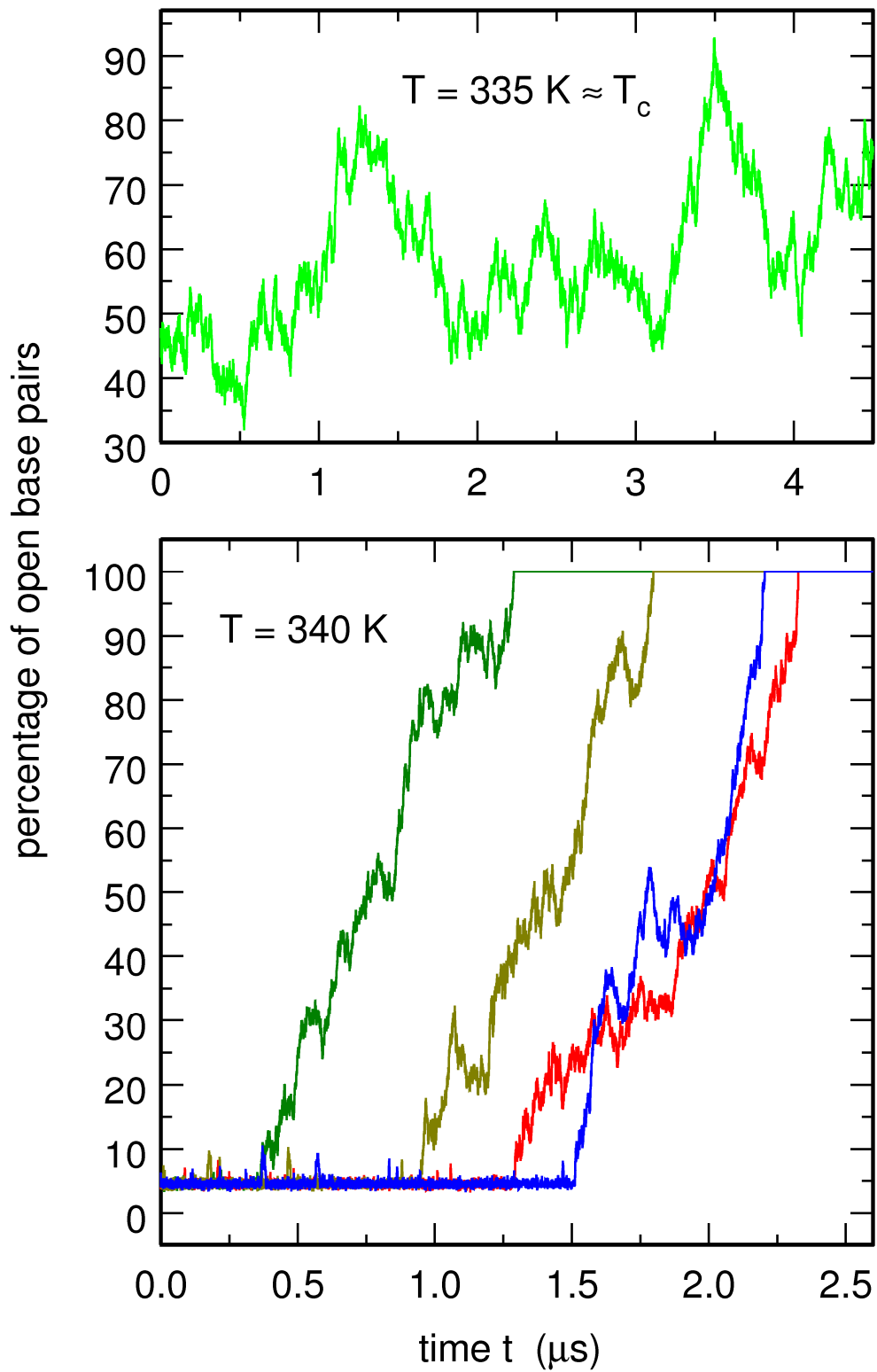

**FIGURE 4**

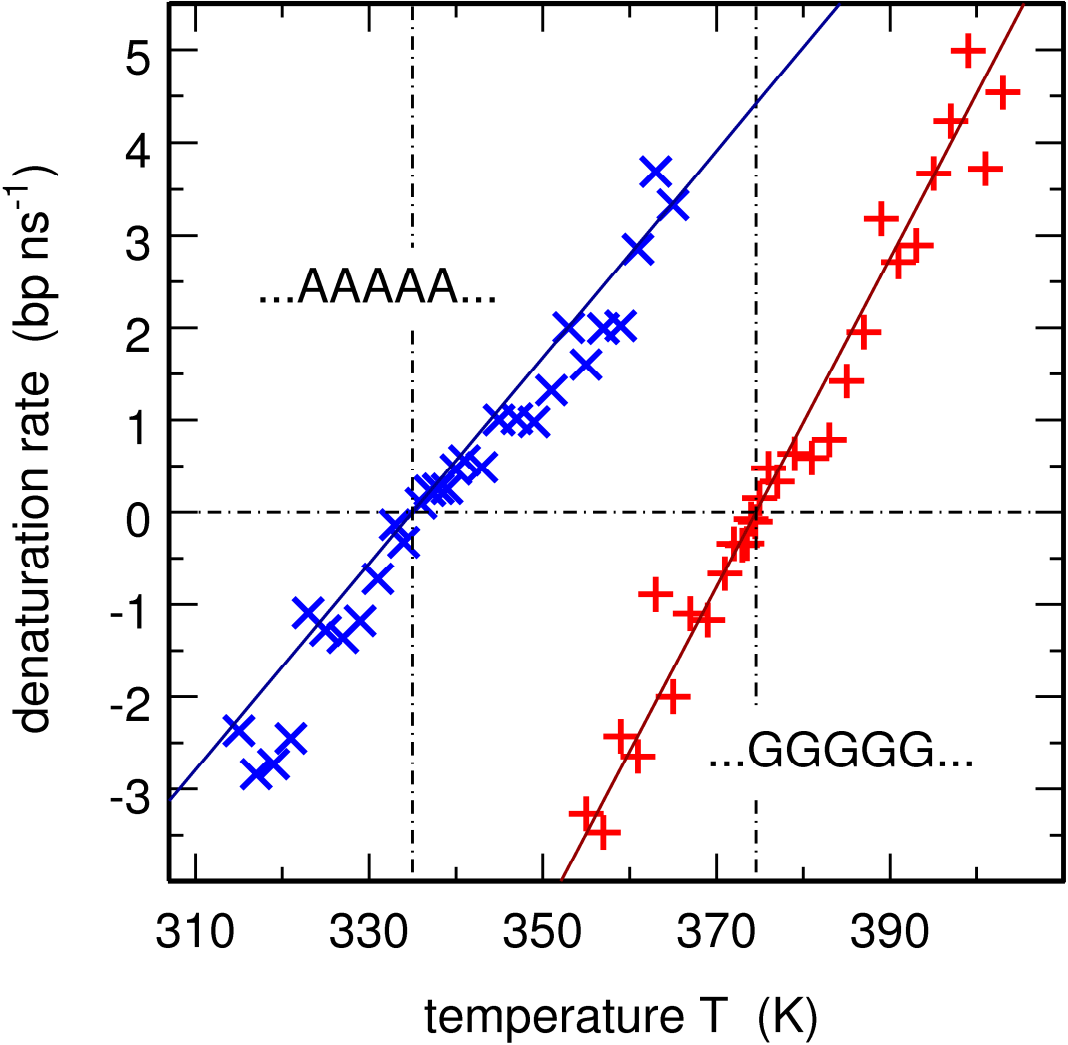

**FIGURE 5**

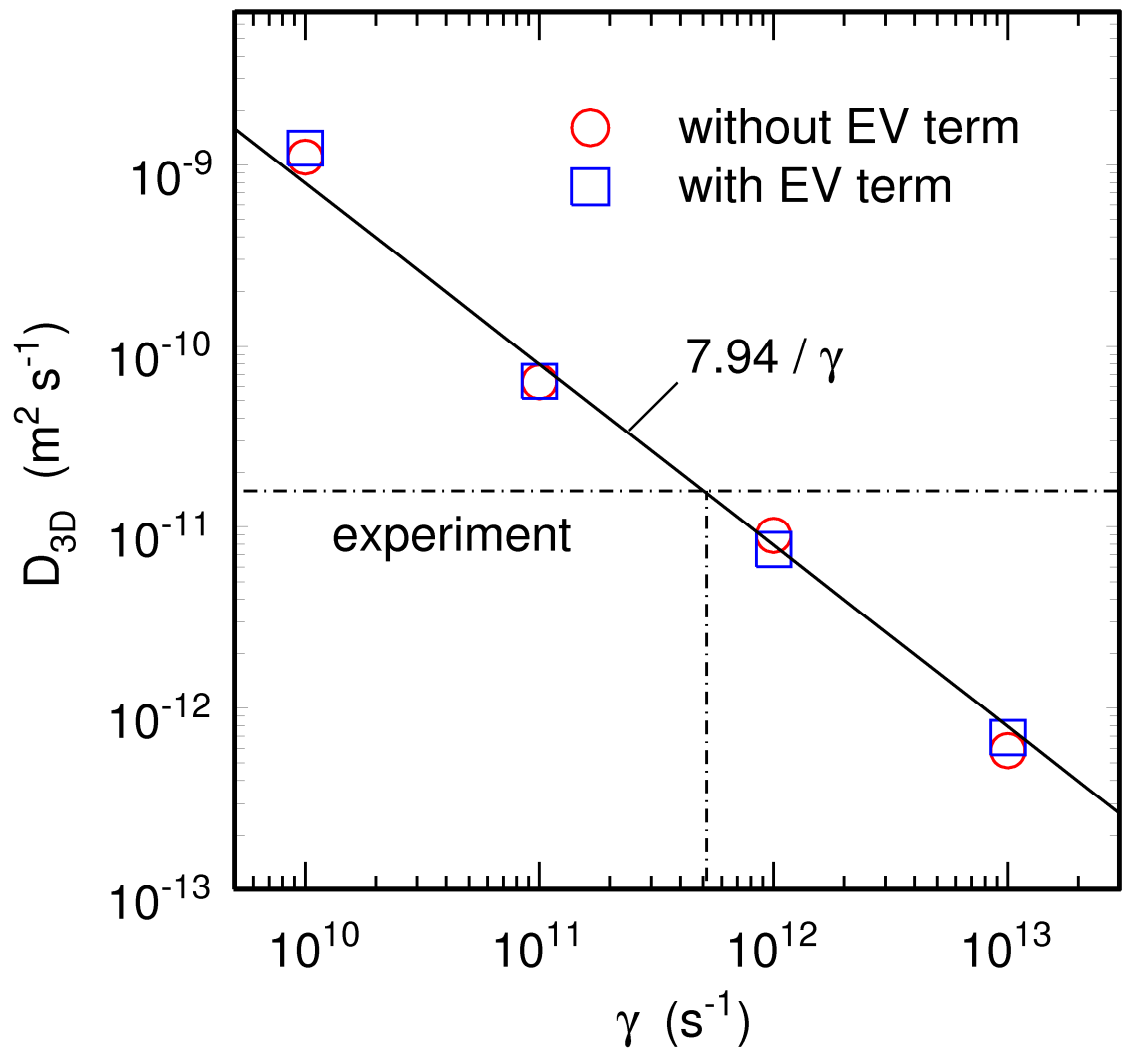

**FIGURE 6**

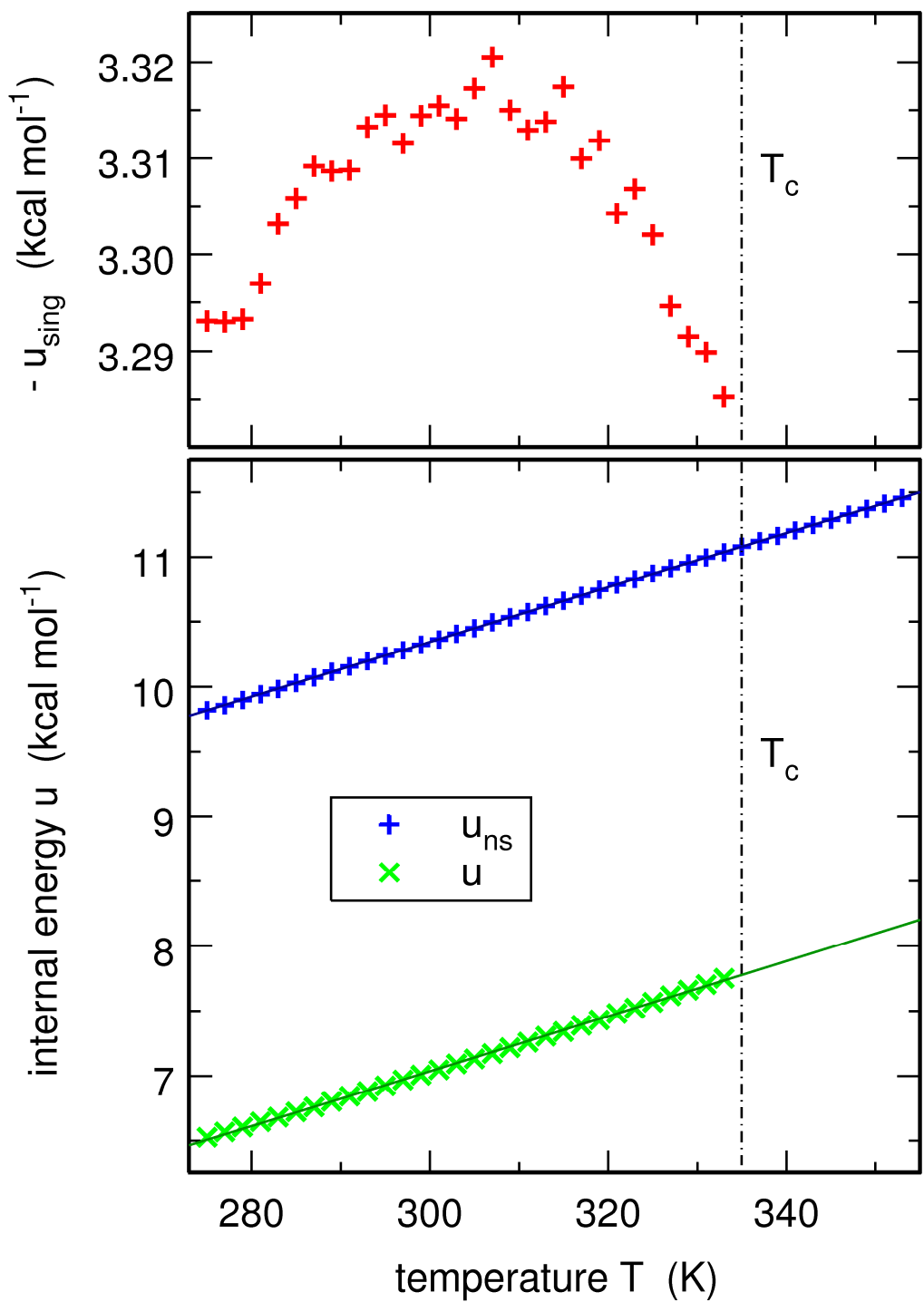

**FIGURE 7**

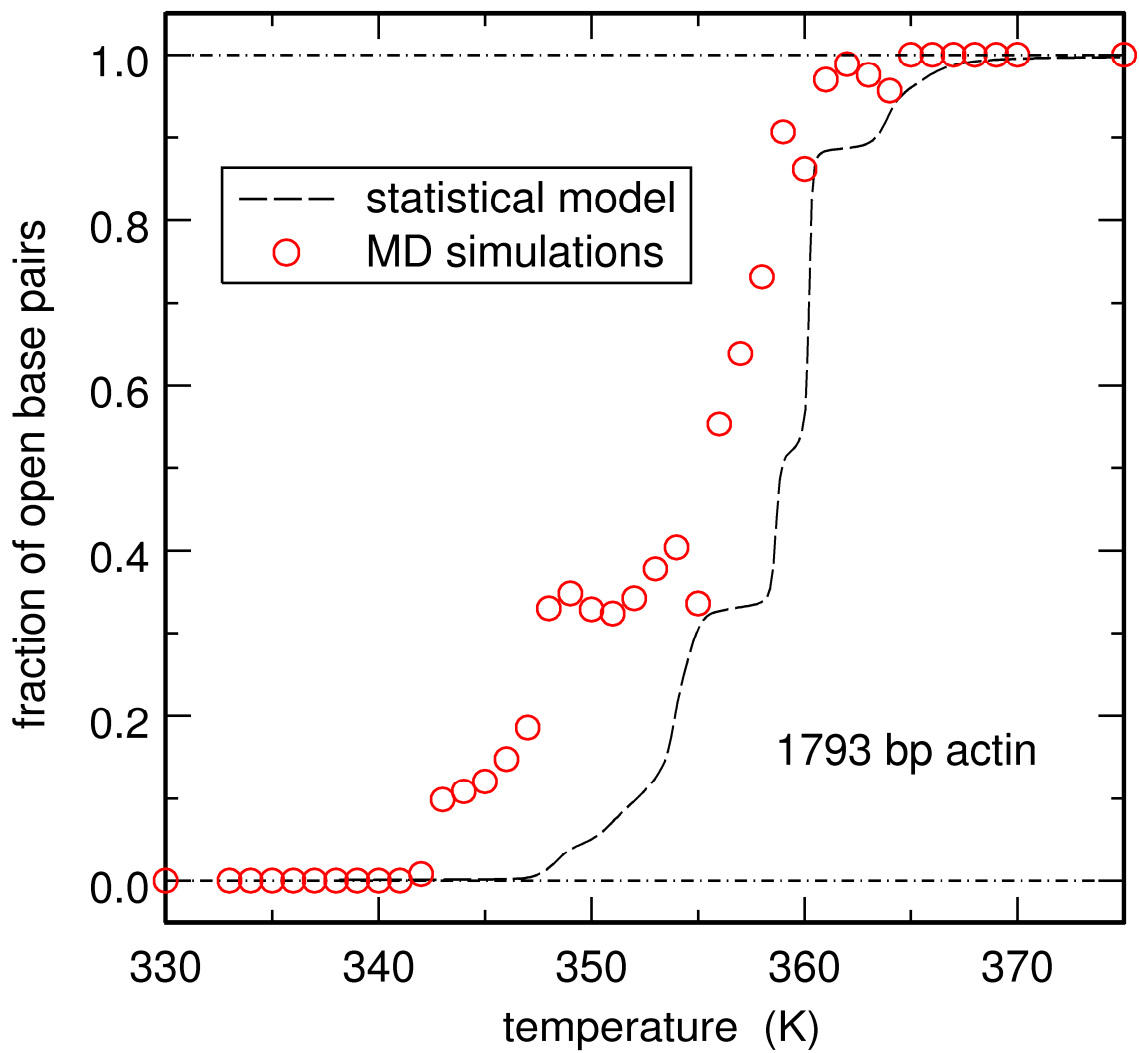

**FIGURE 8**

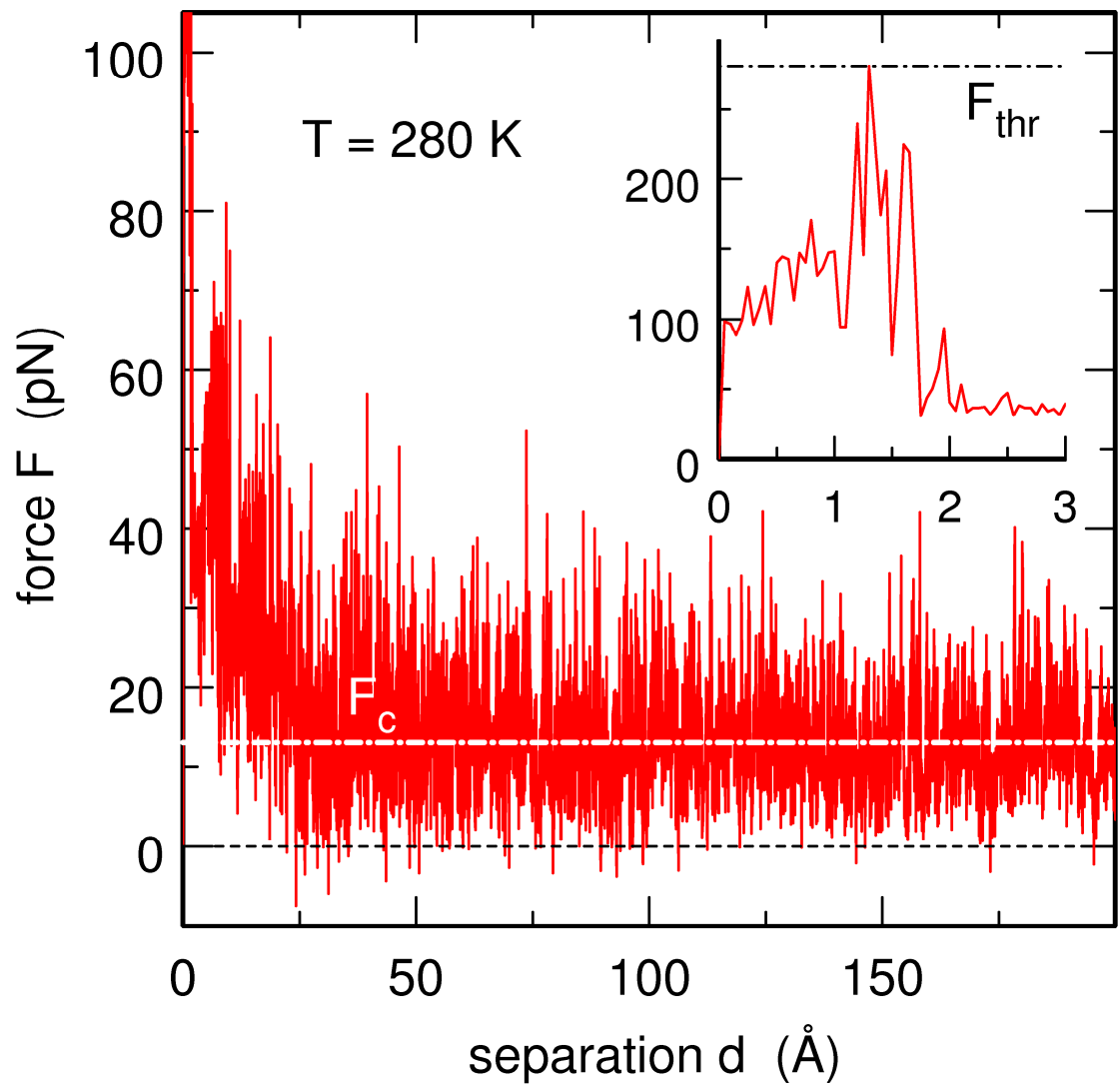

**FIGURE 9**

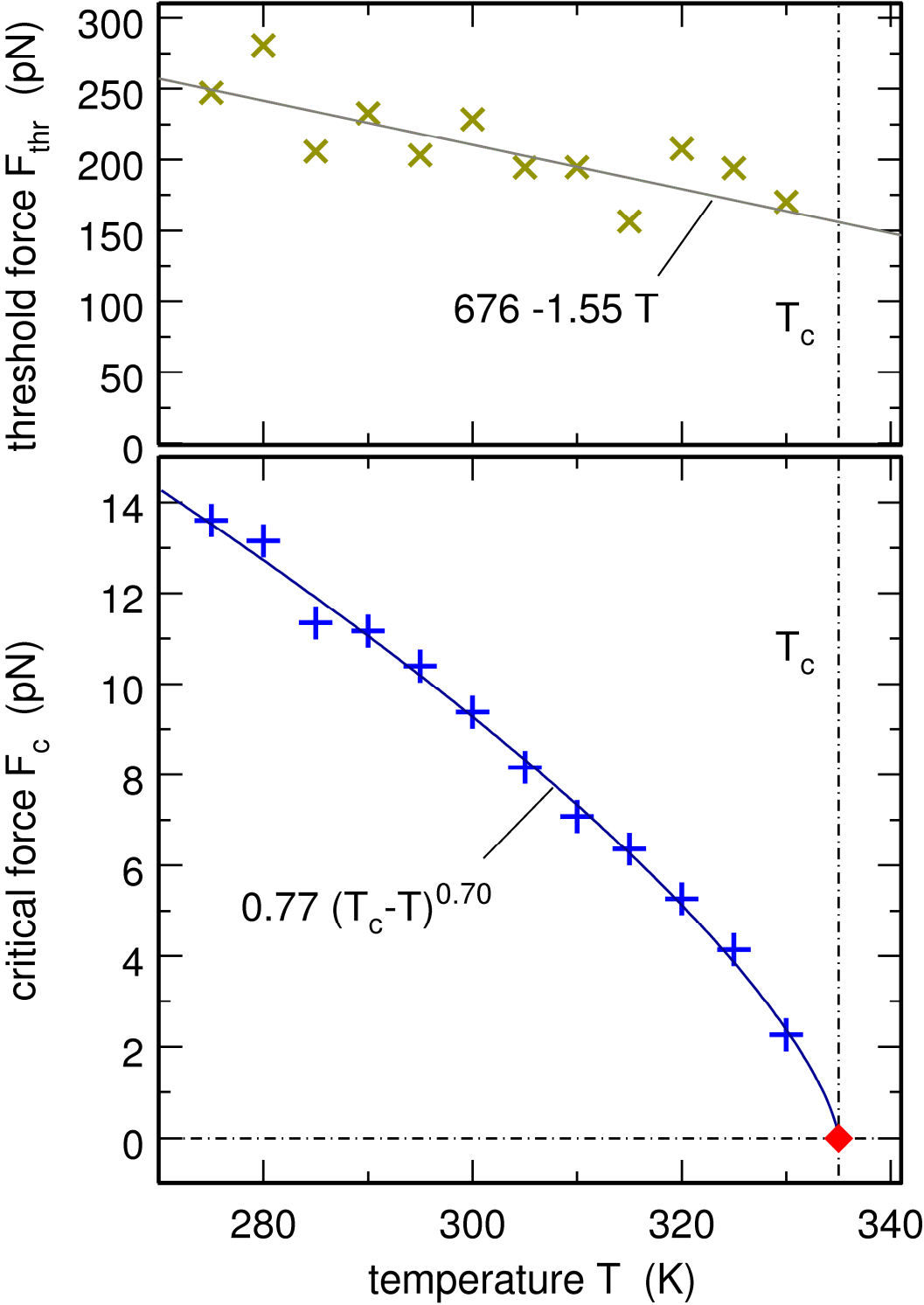

**FIGURE 10**

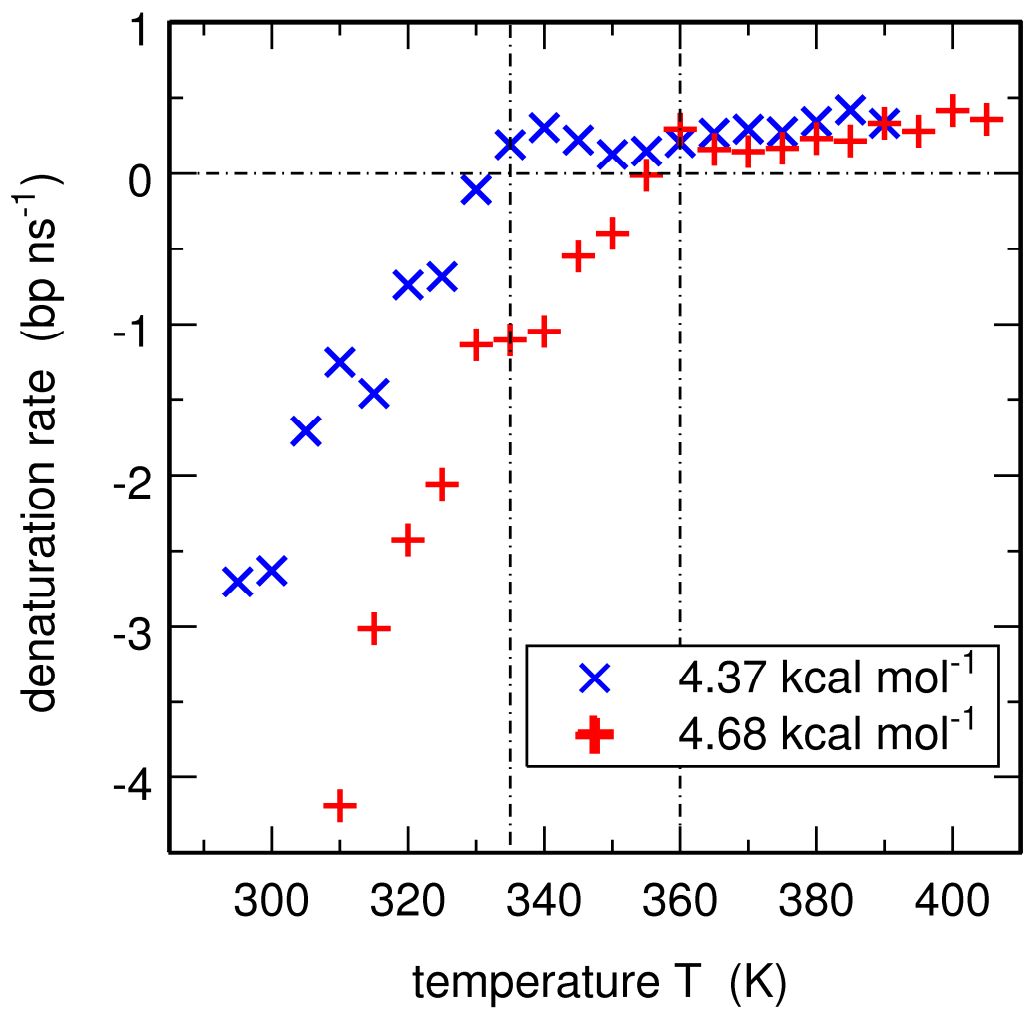

**FIGURE 11**

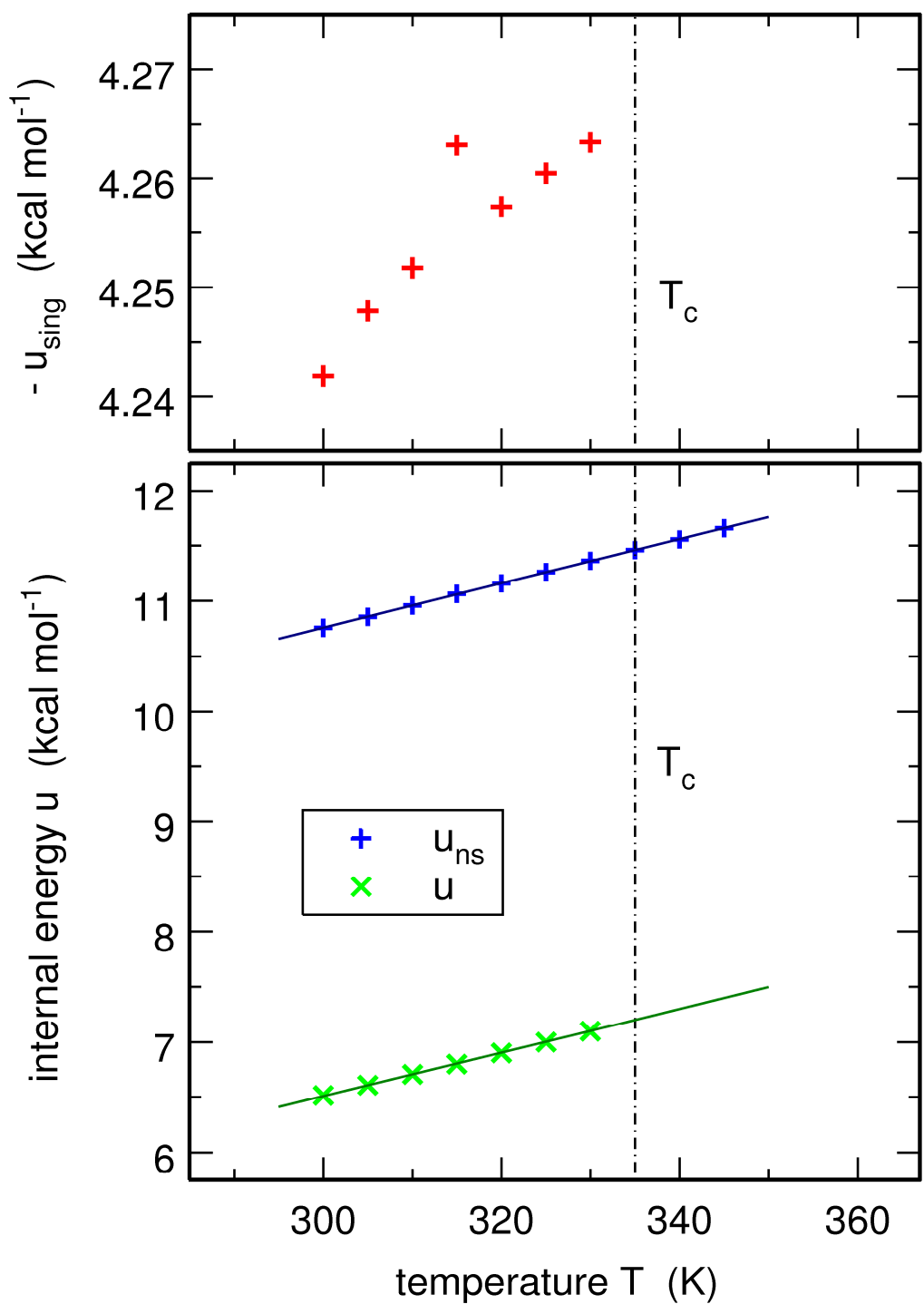

**FIGURE 12**

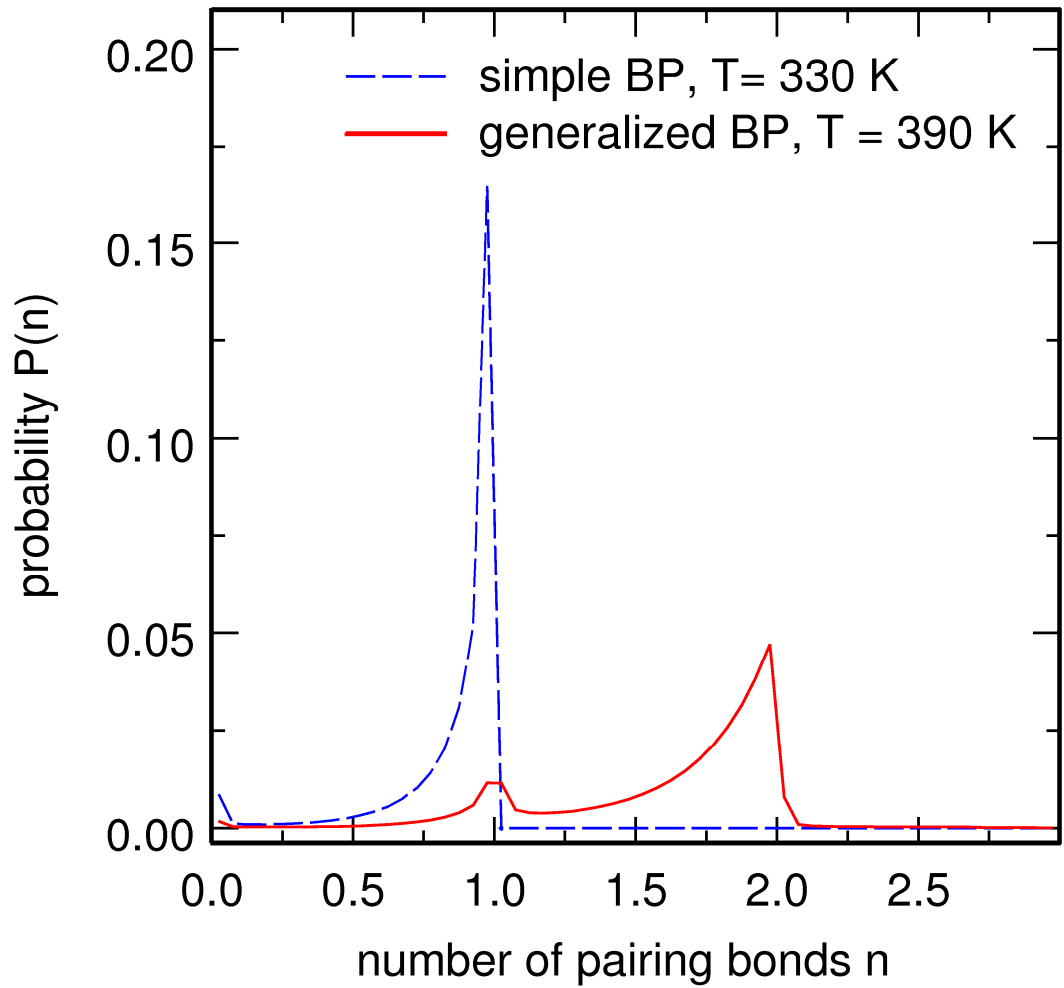

**SUPPLEMENTARY FIGURE S1**

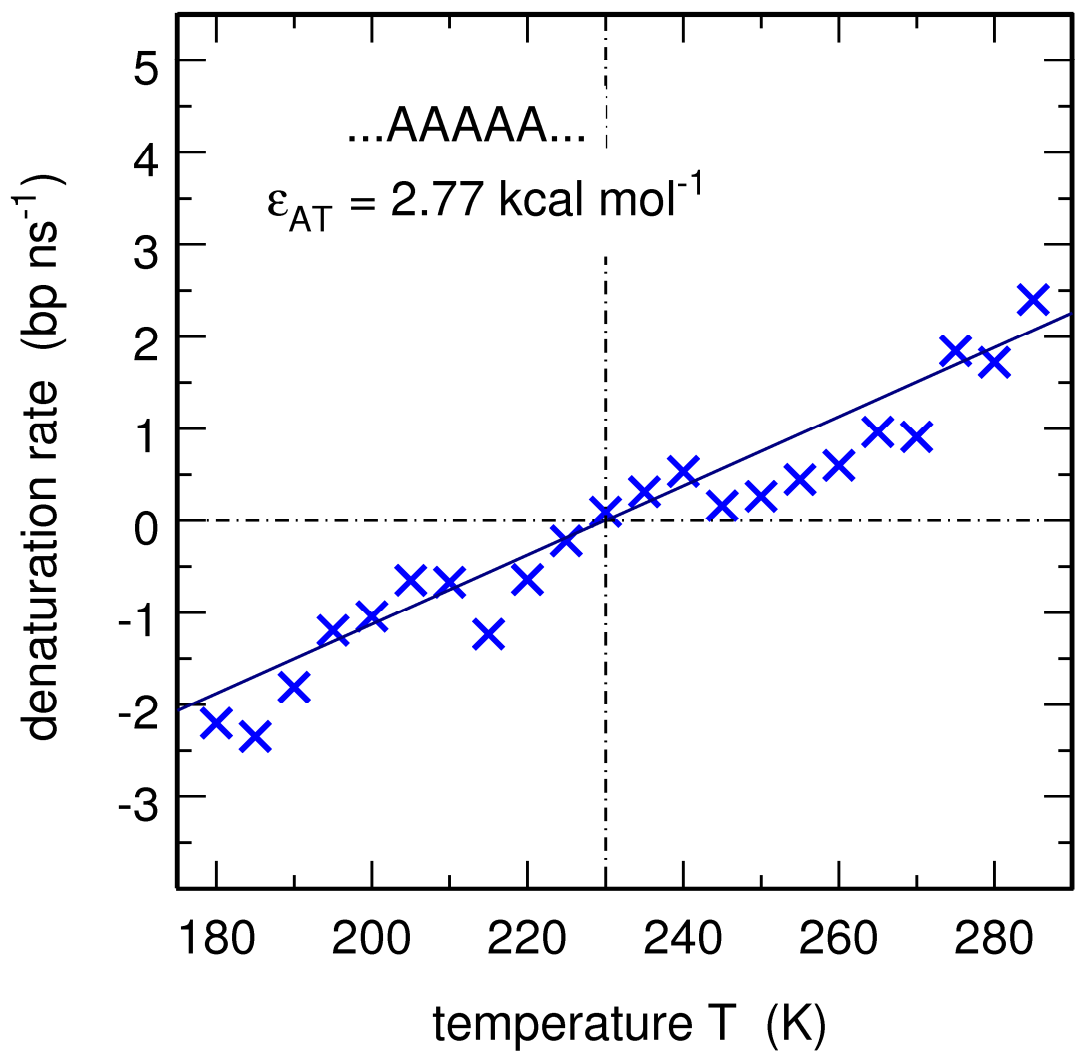

**Figure S1** : Denaturation/annealing rates for a 480 bp ...AAAAA... sequence as a function of temperature, when the base pairing energy of Ref. [78] ($\varepsilon_{AT} = 2.77$ kcal mol$^{-1}$) is assumed instead of that derived in Section II ($\varepsilon_{AT} = 3.90$ kcal mol$^{-1}$). Positive (respectively, negative) rates correspond to denaturation (respectively, annealing). The vertical dash-dotted line indicates the position of the critical temperature (about 230 K) where the rate is zero. These rates were obtained by starting simulations with half-open sequences like the one shown in the left side of Fig. 2 and in checking how long it takes for these sequences to open or close completely. It was furthermore assumed that $V_{bp}$ acts only between bases belonging to the same pair $i$, as for the results displayed in Fig. 4.

**SUPPLEMENTARY FIGURE S2**

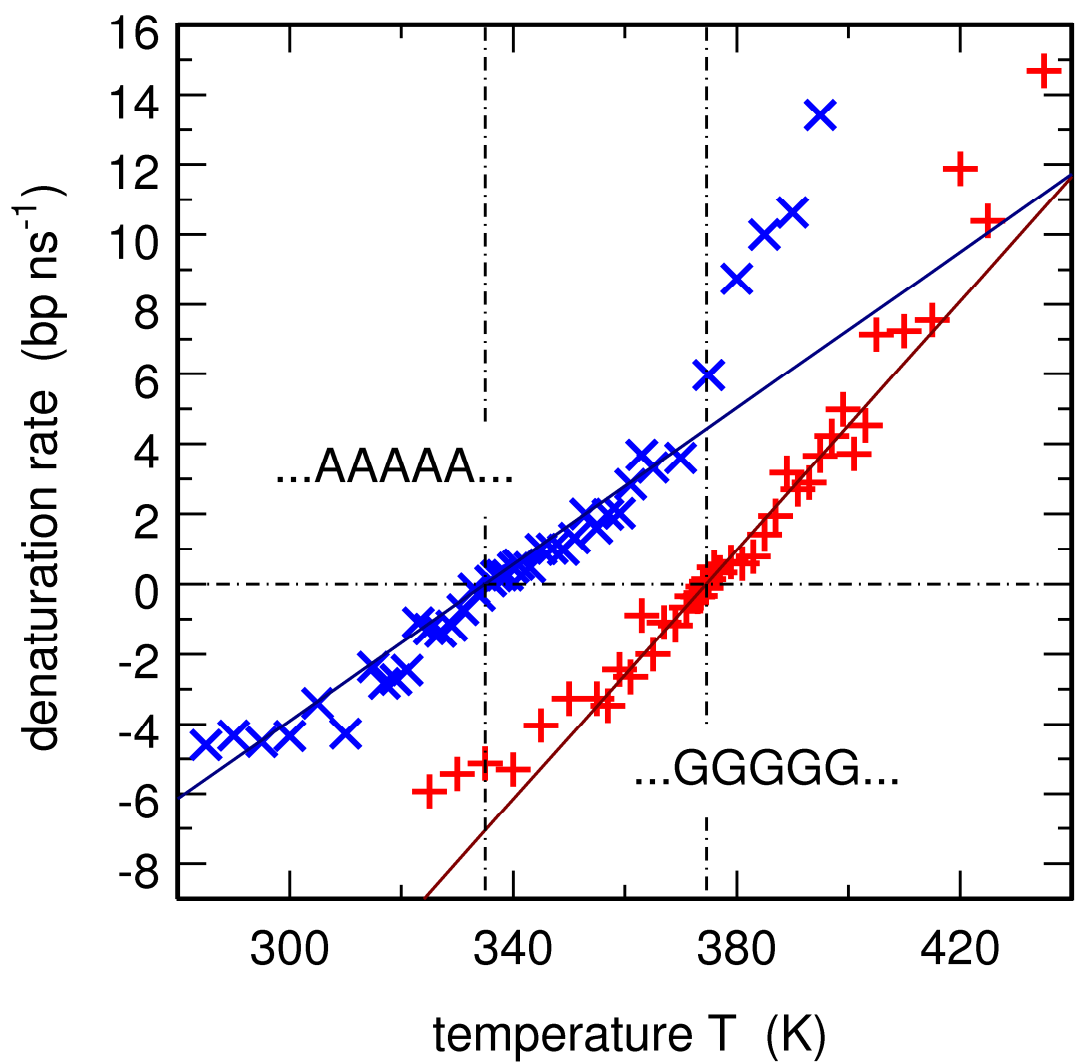

**Figure S2** : Same as Fig. 4, but on broader temperature and rate scales.